\documentclass[%
 reprint,
superscriptaddress,
 amsmath,amssymb,
 aps,
 utf8,
 latin1,
]{revtex4-2}
\pdfoutput=1 
\usepackage{graphicx,caption}% Include figure files
\usepackage{dcolumn}% Align table columns on decimal point
\usepackage{bm}% bold math
\usepackage{gensymb}
\usepackage{indentfirst}
\usepackage{amsmath}
\usepackage{color}
\usepackage{wasysym}
\usepackage{capt-of}
\usepackage{multirow}
\usepackage[utf8]{inputenc}
\usepackage{caption}
\usepackage[flushleft]{threeparttable}
\begin{document}

\preprint{APS/123-QED}

\title{Accurate theoretical determination of the ionization potentials of CaF, SrF, and BaF}% Force line breaks with \\

\author{A. A. Kyuberis}
\affiliation{Van Swinderen Institute for Particle Physics and Gravity, University of Groningen, Nijenborgh 4, 9747AG Groningen, The Netherlands}
\author{L. F. Pa\v{s}teka}
\affiliation{Van Swinderen Institute for Particle Physics and Gravity, University of Groningen, Nijenborgh 4, 9747AG Groningen, The Netherlands}
\affiliation{Department of Physical and Theoretical Chemistry, Faculty of Natural Sciences, Comenius University, Bratislava, Slovakia}
\author{E. Eliav}
\affiliation{School of Chemistry, Tel Aviv University, 6997801 Tel Aviv, Israel}
\author{H. Perrett}
\affiliation{School of Physics and Astronomy, University of Manchester, Oxford Rd, Manchester M13 9PL, UK}
\author{A. Sunaga}
\affiliation{Department of Physics, Graduate School of Science, Kyoto University, Kyoto 606-8502, Japan}
\author{S. M. Udrescu}
\affiliation{Massachusetts Institute of Technology, Cambridge, MA 02139, USA}
\author{S. G. Wilkins}
\affiliation{Massachusetts Institute of Technology, Cambridge, MA 02139, USA}
\author{R.~F.~Garcia Ruiz}
\affiliation{Massachusetts Institute of Technology, Cambridge, MA 02139, USA}
\author{A. Borschevsky}
\affiliation{Van Swinderen Institute for Particle Physics and Gravity, University of Groningen, Nijenborgh 4, 9747AG Groningen, The Netherlands}
\date{\today}% It is always \today, today,

\begin{abstract}
We present a comprehensive theoretical study of the ionization potentials of the MF (M= Ca, Sr, Ba) molecules using the state-of-the-art relativistic coupled cluster approach with single, double, and perturbative triple excitations (CCSD(T)). We have further corrected our results for the higher order excitations (up to full triples) and the QED self energy and vacuum
polarisation contributions. We have performed an extensive investigation of the effect of the various computational parameters on the calculated ionisation potentials, which allowed us to assign realistic uncertainties on our predictions. For CaF and BaF, where precise experiments are available, our predictions are in excellent agreement with the measured values. In case of SrF, we provide a new accurate prediction of the ionisation potential that deviates from the available experimental data, motivating further experimental investigations.

%Ionization potentials play a crucial role in understanding the electronic structure and reactivity of molecules. Accurate determination of these properties is essential for various fields, including chemistry, physics, and materials science.
%Our results reveal that both relativistic effects and basis set size significantly influence the calculated ionization potentials of CaF, SrF, and BaF. Extrapolation to the CBS limit does not seem to be reliable in the framework used for these calculations. Additionally, with increasing of the atomic number influence of QED effects become more important and essential to be taken into account.
%Furthermore, we compare our theoretical results with available experimental data to validate the accuracy of our calculations. 
\end{abstract}

\maketitle

%\tableofcontents

% \section{\label{sec:level1}First-level heading:\protect\\ The line
% break was forced \lowercase{via} \textbackslash\textbackslash}

% The study of radioactive molecules poses significant challenges. The additional vibrational and rotational degrees of freedom massively increases the number of states that can be populated at a given temperature compared to atoms. This is compounded by the small 

\section{Introduction}
Ionization potentials play a crucial role in understanding the behavior of atoms and molecules, thereby serving as a fundamental concept in both chemistry and physics. They provide valuable information about the electronic structure of atoms and molecules, enabling the determination of energy levels and the electronic configurations. Knowledge of the accurate values of ionization potentials contributes to the development and refinement of theoretical models and computational methods and facilitates understanding of quantum phenomena, such as electron correlation, many-body interactions, and the behavior of excited states. 
The ionization potentials of all the naturally occurring elements have been determined experimentally, mostly to very high precision \cite{LinMal17}. Considerable progress in measuring the IPs of the heavier actinides was made in recent years, where No ($Z=102$) \cite{ChhAckBac18} and Lr ($Z=103$) \cite{SatAsaBor15} are the heaviest elements where the IPs were measured using challenging one-atom-at-a-time experimental techniques. Accurate measurements of molecular ionization potentials are much more scarce, even for the simplest molecules, composed of only two atoms. However, such measurements that %The scarcity of accurate measurements of molecular ionization potentials becomes particularly significant. These measurements, even for the simplest diatomic molecules,
provide accurate fundamental information about the molecules hold immense importance in the context of precision experiments on atoms and molecule. Table-top molecular experiments offer a distinct avenue to explore physics beyond the Standard Model and the potential violation of fundamental symmetries, such as time reversal ($\mathcal{T}$) and parity ($\mathcal{P}$) symmetries \cite{SafBud18,Alr22}. 
%Molecules, in particularly dimers are good for precision measurements, including search for BSM physics. These experiments require theoretical support, but many needed properties are not measurable. Hence we need a proxy property, to see how these methods do. Thus, we calculate IPs. 
%___________________________________________________

%Precision experiments on atoms and molecules provide a unique opportunity to search for physics beyond the Standard Model, including the search for violation of fundamental symmetries, such as time reversal ($\mathcal{T}$) and parity ($\mathcal{P}$) symmetries \cite{SafBud18}. 
%There are certain reasons why atoms and molecules have very promising applications to the tests of fundamental symmetries. 
%Spectroscopy with high resolution and high sensitivity has been developed enormously over the past years, enabling the observation of tiny optical effects induced by novel interactions. 
Molecular electronic structure leads to enhancements of the tiny effects of the various $\mathcal{P}$- and $\mathcal{P},\mathcal{T}$-violating phenomena, bringing these into the reach of experimental precision~\cite{SafBud18,Arr23}.
Compared with atoms and polyatomic molecules, diatomic molecules have additional significant advantages, one of which is that calculations of diatomic molecules are not very computationally expensive, allowing highly accurate theoretical investigations.
%These molecules have a high sensitivity to electric fields, making them promising candidates for precision measurements of the EDM, which can provide insights into physics beyond the Standard Mode.
Furthermore, certain diatomic molecules can be laser-cooled to ultracold temperatures \cite{Tar18,FitTar21}, increasing the interaction times and offering  opportunities to explore phenomena such as quantum degeneracy, ultracold chemistry, and quantum simulation \cite{LadTom23,KasKre23,SonPar00}, which have implications for both fundamental physics and quantum information science.
These challenging experiments greatly benefit from reliable theoretical support, from the planning stage to the analysis of the results. In particular, interpretation of experiments searching for violation of fundamental symmetries requires knowledge of various coupling parameters, which are needed to extract the properties of interest (e.g. the magnitude of the electric dipole moment of the electron (eEDM) or the anapole moment of a certain nucleus) from the measured energy shifts \cite{SafBud18, HaaDoe21}. These parameters depend on the molecular electronic structure and cannot be measured, and hence have to be provided through theoretical investigations.

Clearly, such theoretical predictions should be based on accurate and reliable calculations, performed using state-of-the-art computational methods that treat both electron correlation and relativistic effects on a high level. It is also of great importance in an experimental context to provide uncertainty estimates along with any predicted values. Performance of the employed computational tools and the schemes for error estimation should thus be benchmarked against known accurate experimental values, to explore their reliability and predictive power for systems and properties where no experiment is available. 

Relativistic coupled cluster approach is considered to be one of the most powerful tools for treatment of heavy many-electron systems. It has demonstrated high accuracy and strong predictive power for a variety of atoms and molecules and for different properties \cite{EliBorKal14,EliBorZai22}. Recently, we have developed a scheme that allows us to set error bars on the calculated properties, by performing an extensive study of the effect of the different computational parameters on the results \cite{LeiKarGuo20,HaaEliIli20}. 

In this work, we apply the relativistic coupled cluster approach to the alkaline earth metal fluoride molecules, MF, where M = Ca, Sr and Ba. Group II fluorides benefit from a relatively simple electronic structure that allows for their laser cooling~\cite{ShuBar10,ZheCou14,KogRocAlb21} and facilitates the interpretation of the experimental results. BaF is considered for precision measurements aimed to set a new upper limit on the electric dipole moment of the electron \cite{AggBet18} and to measure the anapole moment of the $^{137}$Ba nucleus \cite{AltAmmCah18}. Such experiments require input from theory for their interpretation \cite{HaoIliEli18,HaaDoe21} and benchmarking of calculations on these systems for properties where experiment is available is crucial. We thus apply our method of choice to theoretical predictions of ionization potentials of the MF molecules, including uncertainty estimates.

%Firstly, Group II fluorides have a relatively simple electronic structure, which simplifies the interpretation of experimental results. The outermost electron configuration of these molecules consists of a single valence electron, making the behavior of this electron easier to comprehend and analyze. This simplicity facilitates the precise measurement and interpretation of potential shifts induced by an electric field, enabling researchers to study and quantify, for example, the eEDM effects more effectively.
%Additionally, Group II fluorides can be laser-cooled to extremely low temperatures. Laser cooling enhances the sensitivity of eEDM measurements by reducing thermal motion-related uncertainties, improving the precision and accuracy of the obtained results.
%Overall, the combination of a straightforward electronic structure and the ability to laser cool Group II fluorides, such as CaF, SrF, and BaF, provides researchers with a robust platform for investigating electric dipole moments. 
%These advantages contribute to a deeper understanding of fundamental physics and may potentially lead to the discovery of new physics beyond the standard model.

Numerous theoretical studies of the electronic structure and other properties of CaF, SrF, and BaF were carried out using a variety of methods~\cite{BaFp,KanKua16,TohKor15,PhysRevA.89.053416,10.1063/1.447394,HaaEliIli20}. Experimental investigations were mainly dedicated to measuring various spectroscopic properties of these molecules, ~\cite{LinMal17,BarBea67,DagCruZar03,FieHar75,SteDomHar77,NakDomSte78,SheWanDic09,NitSchErn88,ColGuoZha96,KalBloMcc99,BerOlsCha93}, such as molecular dipole moments and suitability for laser-cooling~\cite{ShuBar10,ZheCou14}. The ionization potentials of CaF and BaF were measured in 1990s, by methods of Rydberg spectroscopy \cite{MurBerMer90,Jakubek1995}, and much an earlier value is available for SrF \cite{Hildenbrand1968} obtained from a pure electron-impact measurement.
The aim of this work is thus twofold: to benchmark the quality of our calculations against accurate experimental values in the case of CaF and BaF and to provide an accurate and reliable prediction for SrF, where only a low precision experiment is available. 
In the following, Section \ref{comp-meth} presents the methods and computational details employed in this work. %employed in this paper are introduced. 
Section \ref{results} contains the calculated values while in Section \ref{uncer} the procedure for uncertainty evaluation used in this work is outlined. 
The final recommended values, along with comparison to earlier calculations and experiments, are presented in Section \ref{fin-res}. %those are recommended to use, for calculated IP values of MF (M = Ca, Sr, Ba) with uncertainties.

%The IP of the group II monofluorides decreases progressively with increasing proton number, before increasing again in RaF.

\section{Method and computational details}\label{comp-meth}

All the calculations in this work were carried out using the development version of the DIRAC code \cite{dirprog,SunSalSau22}, which allows us to use relativistic methods, in particular the  traditional 4-component Dirac-Coulomb (DC) Hamiltonian.
%precise version and build information is found in the output files, see \cite{SunSalSau22}. 

The single reference coupled cluster approach with single, double (CCSD), and perturbative triple excitations (CCSD(T)) was used in all the calculations. 
%In the most CCSD(T) calculations the virtual space cut off was established at the level of 50 a.u.. \\

In order to take full advantage of this state-of-the-art approach, we used the relativistic Dyall basis sets \cite{dyall1,dyall2}. These basis sets are available in different quality, v$n$z  ($n$ = 2, 3, 4). Furthermore, one can  employ either the “valence” (dyall.v$n$z), the “core-valence” (dyall.cv$n$z), or the "all-electron" (dyall.ae$n$z) variants of the basis sets of the same principal number; the latter 2 include extra core-valence- and core-correlating functions, respectively. Finally, extra layers of diffuse functions can be added, designated by $k$-aug-v$n$z, with $k$ = s (single), d (double), etc. Presence of diffuse functions improves the description of the bond region and hence of the valence properties. 

%The calculated properties were extrapolated to the complete basis set limit (CBSL), based on the $s$-aug-cv$n$z results, using using the scheme of Feller \textit{et al.} \cite{Feller1992} for the DHF energy and that of Helgaker \textit{et al.} \cite{cbs} for the correlation contribution.  
%The valence basis sets are defined to include functions for the correlation of the outer ns shell and the (n-1)s and p shells for the s elements, the outer ns and np shells for the p elements, the ns, np, nd, and (n+1)s for the d elements, and the ns, np, nd, nf, (n+1)s, (n+1)p, (n+1)d, and (n+2)s for the f elements. The core-valence basis sets include the (n-2) shell for the s elements, the (n-1) shell for the p elements, the (n-1) shell for the d elements, and nothing extra for the f elements.

%For IP calculations and subsequent uncertainty evaluation we would do following procedure. 

The following procedure was employed for the calculations of the molecular ionization potentials. We calculated the potential energy curves of both the neutral molecule and the molecular ion of MF (M = Ca, Sr, Ba) using the DC-CCSD(T) approach and different basis sets. The ionization potential was then obtained as the difference between the energies of MF and MF$^+$ at the corresponding equilibrium bond lengths. The final results were obtained from a complete basis set limit (CBSL) extrapolation of the potential energy curves based on the the s-aug-c$\textit{n}$vz ($n=2,3,4$) basis sets and carried out  following the scheme of Helgaker et al. \cite{cbs} (H-CBSL). 
While this scheme is one of the most popular CBSL extrapolation approaches, we also test two other possibilities: the scheme of Martin \cite{cbs1} and the scheme of Lesiuk and Jeziorski \cite{cbs2}. These results are consistent to within 1 meV (with the H-CBSL in between the two other values), confirming the convergence of the calculated IP with the respect to the basis set cardinality.
Using these results the adiabatic IPs and spectroscopic molecular properties were extracted.%, see Tables ~\ref{constants} and ~\ref{cbs-sch}. 

\section{Results}\label{results}

Table ~\ref{constants} contains the calculated spectroscopic constants of the ground states of the neutral molecules and molecular ions, compared to earlier theoretical predictions and experimental values, where available. In these calculations we used the DC-CCSD(T) approach combined with the CBS extrapolation of the basis sets. 25, 35 and 35 electrons were correlated for CaF, SrF, and BaF, respectively, and the virtual space cutoff was set at 50 a.u. Our results are in a good agreement with previous calculations and experimental values. The least numerically stable value is anharmonicity correction $\omega_{e}\chi_{e}$. However, for these constants the experimental uncertainty is often rather high. 

  \begin{table}[h]
        \centering
        \caption{Spectroscopic constants of the ground state of MF (M = Ca, Sr, Ba) and corresponding ions, calculated on the DC-CCSD(T) level. Results are based on CBS extrapolation limit and 25, 35 and 35 electrons were correlated, respectively, with the virtual space cutoff at 50 a.u. $\omega_e$, $\omega_{e}\chi_{e}$ are shown in cm$^{-1}$.}\label{constants}
        \begin{tabular}{cccccc}
        \hline
Molecule&$R_e$(\AA)&$\omega_e$&$\omega_{e}\chi_{e}$&Ref.&Method\\
              \hline
CaF&1.953&588.1&3.03&Present&CCSD(T)\\
   &1.958&586.2&2.90&\cite{BaFp}&FSCC\\
   &2.001&572.4&2.70&\cite{YANG2007147}&MRCI\\
   &1.971&612.5&3.70&\cite{PelVivCir05}&MRCI\\
   & 1.967&--& -- &\cite{LinMal17}&Experiment\\
   & --& 581.1& -- &\cite{FieHar75}&Experiment\\
   
CaF$^{+}$&1.871&693.5&2.78&Present&CCSD(T)\\
\hline
SrF&2.076&501.1&1.92&Present&CCSD(T)\\
   &2.083&500.1&2.45&\cite{BaFp}&FSCC\\
   &2.137&475.0&--&\cite{JarKorYou14}&CASSCF+MRCI\\
   &2.076&--&--&\cite{BarBea67}&Experiment\\
   &--&502.4&2.27&\cite{SteDomHar77}&Experiment\\
   &--&502.0&2.21&\cite{ColGuoZha96}&Experiment\\
   &--&502.4&2.20&\cite{DomSteHar77}&Experiment\\

SrF$^{+}$&1.996&587.9&2.16&Present&CCSD(T)\\
\hline
BaF&2.162&467.6&2.32&Present&CCSD(T)\\
    &2.177&468.4&1.83&\cite{BaFp}&FSCC\\ 
    &2.171&474.1&1.90&\cite{KanKua16}&CASSCF+MRCI+SOC\\
    &2.159&--&--&\cite{RyzTor80}&Experiment\\
    &--&469.4&--&\cite{BarBer88}&Experiment\\
    &--&--&1.84&\cite{BerEff92}&Experiment\\
%      &-&468.9&1.79&\cite{KniEasWel1971}&\\
      
BaF$^{+}$& 2.087&538.3&1.92&Present&CCSD(T)\\
              \hline
        \end{tabular}
    \end{table} 

In Table~\ref{cbs-sch} the IP values obtained using different basis sets are presented. These calculations were carried out correlating 25, 35 and 35 electrons, respectively, and setting the virtual space cutoff at 50 a.u..  

 \begin{table}[h]
        \centering
        \caption{Adiabatic IPs (eV) of MF (M = Ca, Sr, Ba) obtained using basis sets of differing quality. The results were obtained within the CCSD(T) approach, correlating 25, 35 and 35 electrons, respectively, and including virtual orbitals up to 50 a.u.% \textcolor{red}{[Can you explain the large discrepancy in the s-aug-dyall.cv4z and d-aug-dyall.cv4z for CaF?
%In CaF, the difference between "s-aug" and "d-aug" is larger than that of "dyall.cv4z" and "s-aug". The trends of SrF and BaF are reasonable for me.]}
}\label{cbs-sch}
        \begin{tabular}{lcccc}
        \hline
              Basis set &CaF&SrF&BaF \\
              \hline
              dyall.v3z&5.8897&5.3949&4.8072\\
              dyall.cv3z&5.8257&5.4014&4.8043\\
              dyall.ae3z&5.8257&5.4014&4.7967\\
              dyall.cv4z&5.8199&5.4075&4.7991\\
%              s-aug-dyall.cv3z&5.843&5.415&4.808\\
              s-aug-dyall.cv4z&5.8280&5.4152&4.8026\\
%              d-aug-dyall.cv3z &5.843 &5.416&4.809\\
              d-aug-dyall.cv4z & 5.8266 &5.4155&4.8030\\
              s-aug-dyall.cv$n$z H-CBSL&5.8236 &5.4174&4.8035\\
%              s-aug-dyall.cvnz  LJ-CBSL &5.823&5.418&4.804\\
%               s-aug-dyall.cvnz  M-CBSL& 5.824&5.417&4.803\\
        \hline
        \end{tabular}
    \end{table} 
    
The adiabatic IPs, derived from the molecular energies calculated at the equilibrium bond lengths of the neutral molecules and the molecular ions, were further corrected for the higher order correlation and relativistic contributions. 

To correct for the limited active space used, we calculated the difference between the results obtained correlating electrons occupying orbitals with energies above --20 a.u. and a virtual space cutoff of 50 a.u. and those obtained correlating all electrons (29, 47, and 65, for CaF, SrF, and BaF, respectively) with a corresponding virtual space cutoff of 2000 a.u. In order to capture the full active space effect and to account for inner-core correlations, the all-electron quality basis sets were used in the latter calculations. For these calculations, we used the dyall-cv3z and dyall-ae3z basis sets. The results are shown in the Table~\ref{active_space}. 
%I put now 3z, but for journal submission 4z results will be ready, they are still running%

 \begin{table}[h]
        \centering
        \caption{Adiabatic IPs (eV) of MF (M = Ca, Sr, Ba), limited active space is orbitals with energies above --20 a.u. and a virtual space cutoff of 50 a.u.}\label{active_space}
        \begin{tabular}{lcccc}
        \hline
              Basis set &active space&CaF&SrF&BaF \\
              \hline
          dyall.cv3z&limited active space&5.8257&5.4014&4.8043\\
          dyall.cv3z&full active space&5.8258&5.4015&4.8052\\
          dyall.ae3z&full active space&5.8258&5.4015&4.7978\\
        \hline
        \end{tabular}
    \end{table}

   \begin{table}
        \centering
        \caption{Calculated IPs of MF (M = Ca,Sr,Ba) [eV] including higher order contributions.}\label{Corrections}
        \begin{tabular}{lccc}
        \hline
    Method & CaF &SrF &BaF \\
     \hline 
    DC-CCSD&\phantom{--}5.8180&\phantom{--}5.3940&\phantom{--}4.7700\\
    DC-CCSD(T) &\phantom{--}5.8236 &\phantom{--}5.4174&\phantom{--}4.8036\\
    +augmentation&--0.0015&\phantom{--}0.0003&\phantom{--}0.0004\\
    +core corr.+active space  &--0.0001&--0.0001&\phantom{--}0.0065\\
    +$\Delta$T &\phantom{--}0.0005 &\phantom{--}0.0008&\phantom{--}0.0010\\
    +Breit &--0.0010&--0.0011&--0.0009\\
    +QED &--0.0008&--0.0020&--0.0031\\
        \hline
       Final result &\phantom{--}5.8208&\phantom{--}5.4151&\phantom{--}4.8075\\
             \hline
        \end{tabular}
    \end{table} 

We  evaluated the effect of the residual triple excitations (beyond (T)) by comparing the IPs calculated at the CCSDT and CCSD(T) levels of theory using the MRCC code \cite{MRCC1}. These calculations were performed using the dyall.v3z basis sets, correlating 15, 17 and 17 electrons for MF, M = Ca, Sr, Ba, respectively, and with a virtual space cutoff set at 10 a.u. As the effect was very small (about 1 meV) we did not consider even higher excitations. 

%The calculations in this work were performed in the framework of the DC relativistic Hamiltonian, which does not include the Breit contribution nor the higher order QED corrections. Calculations of these effects for many-electron molecules are not straightforward, though a number of recent implementations have been reported \cite{Skripnikov2021,SunSalSau22}.

QED corrections were calculated using the development version of the DIRAC program package \cite{SunSalSau22}. The Uehling potential \cite{Ueh35} was employed for the vacuum polarization and the effective potential of Flambaum and Ginges for the electron self-energy~\cite{FlaGin05}. We added effective QED potentials to Ca, Sr and Ba nucleus.  
The electronic structure of the MF molecules is very similar to that of M$^+$, and the valence electron is removed from a non-bonding atomic-like orbital. Thus, we calculated the effect of the Breit contribution on the IP of M$^+$. The Breit calculations were performed within the Fock-space coupled cluster approach (FSCC), using the Tel Aviv atomic computational package~\cite{TRAFS-3C}.

Finally, we correct our results for the effect of missing augmentation functions by taking the difference between the s-aug-dyall.cv4z and the d-aug-dyall.cv4z values (Table \ref{cbs-sch}). 

All the corrections were calculated at the equilibrium geometries of neutrals and ions of MF molecules and added to our baseline IPs from Table \ref{cbs-sch}, obtained at the CBS limit. The individual contributions can be seen in Table~\ref{Corrections}.

\section{Uncertainty estimates}\label{uncer}

Reliable estimates of the uncertainty of theoretical predictions of molecular properties is crucial for the support and the interpretations of the experiments that aim to investigate the systems in question. These uncertainty estimates can be obtained from extensive computational investigations; we have successfully employed such procedures in the past for various atomic and molecular properties \cite{prop1,prop2,prop3,LeiKarGuo20}. 

The three main sources of uncertainty in our calculations are the incompleteness of the employed basis set, the approximations in the treatment of the electron correlation, and the missing relativistic effects. As we are considering higher-order effects, we assume these sources of error to be largely independent. In the following each of these contributions will be presented and evaluated separately; the individual contributions to the total uncertainty are given in Table~\ref{uncertainty-all}.

\paragraph{Basis set.}\label{bst}
To evaluate the basis set incompleteness error, we consider the basis set cardinality and the convergence in terms of core-correlating functions and diffuse functions. 
The final results were obtained at the CBS level. We evaluate the cardinality incompleteness error as half of the difference between CBS and s-aug-dyall.cv4z basis set results. The effect of using a limited number of core correlating function is considered in the next section.  
The remaining uncertainty due to the possible lack of additional diffuse functions is evaluated as the difference between the results obtained using the doubly augmented and the singly augmented dyall.cv4z basis sets. 
 
\paragraph{Electron correlation.}\label{ctt}
We consider separately the effect of using a limited active space (virtual space cutoff) and the effect of excitations beyond perturbative triples.
To account for the limited active space, we take the difference between results obtained with a virtual cutoff of 50 a.u. and 2000 a.u. at the dyall.cv3z basis set level correlating 25, 35 and 35 electrons for MF, where M = Ca, Sr, Ba, respectively.
%, and with a virtual cutoff of 50 a.u. and those obtained correlating all electrons and with a virtual space cutoff of 2000 a.u..
%In order to capture the full active space effect and to take into account also the effect of inner-core correlating functions, the all-electron quality basis set was used in the latter calculation. On the other hand, to conserve computational effort, we employed the modest sized dyall-cv4z and dyall-ae4z basis sets.
We take the contribution of the residual triple excitations ($\Delta$T) as the estimate of the error due to the neglect of the higher order excitations. 

\paragraph{Relativistic effects.}\label{rtt}
We assume that the higher order QED contributions are smaller than the calculated Lamb shift and take the latter as the conservative uncertainty estimate. The Uehling and Flambaum-Ginges potentials can include the lowest QED contribution with $Z \alpha^2$, where $\alpha$ is the fine structure constant. And the next order should be $Z^3 \alpha^4$ from the expansion of the bound-state propagator \cite{SunSalSau22}. From this, we may estimate the contribution from the next order by substituting the nuclear charge $Z$ for Ca, Sr, and Ba, and taking the ratio.
%The Breit operator is not available in the DIRAC program package, and thus inclusion of this term in molecular calculations is problematic. 
%\red{ However,  the electronic structure of MF (M = Ca, Sr, Ba) is very similar to that of M$^+$, and the valence electron is removed from a non-bonding atomic-like orbital. Thus, we calculated the effect of the Breit contribution on the IP of M$^+$. These calculations were performed within the Fock-space coupled cluster approach (DCB-FSCC), using the Tel Aviv atomic computational package \cite{TRAFS-3C}.
%We further incorporated the QED correction for the IP of M$^+$ in the form of the model Lamb shift operator of Shabaev \textit{et al.} \cite{ShaTupYer15}. This model Hamiltonian includes the Uehling potential and an approximate Wichmann--Kroll term for the vacuum polarisation (VP) potential \cite{BLOMQVIST197295} and local and nonlocal operators for the self-energy (SE), the cross terms (SEVP), and the higher-order QED terms.}

The magnitude of the various effects contributing to the uncertainty are given in the Table~\ref{uncertainty-all} and the total uncertainty is obtained by combining all the above terms and assuming them to be independent. 

    \begin{table}[h]
        \centering
        \caption{Main sources of uncertainty in the calculated IP of MF (M = Ca, Sr, Ba) in [meV].}\label{uncertainty-all}
        \begin{tabular}{llccc}
        \hline
              Category & Error source& CaF&SrF& BaF \\
              \hline
        Basis set & Cardinality&2.21&1.10 &0.50 \\
                  & augmentation&1.47&0.27&0.37\\
                  & core correlation&0.10&0.10& 6.50\\
        Correlation& virtual space &1.23 &0.15&0.44\\ 
                   & higher excitations&0.5&0.8&1.0\\
        
        Relativity & QED & 0.02&0.15&0.52\\
            \hline
            Total uncertainty, meV &&2.97&1.41&3.06\\
        \hline
        \end{tabular}
    \end{table}

\section{Final Values}\label{fin-res}

The final recommended adiabatic IPs of MF (M = Ca, Sr, Ba), along with the corresponding uncertainties, are presented in Table~\ref{IPfinal}. 
Furthermore, we also present the vibrationally corrected adiabatic IP$_\mathrm{vc}$, defined in the following way:  %\ref{IP-theo}.
%$$
%IP_{vc} = (E^{MF^+}+\frac{1}{2} %\omega_e^{MF^{+}})-(E^{MF}+\frac{1}{2} %\omega_e^{MF}),
%$$
\begin{eqnarray}
 \mathrm{IP}_\mathrm{vc} =(E^{\mathrm{MF^{+}}}+\frac{1}{2} \omega_e^{\mathrm{MF^{+}}}-\frac{1}{4} {\omega_{e}\chi_{e}}^{(\mathrm{MF^{+}})})-\\
 (E^{\mathrm{MF}}+\frac{1}{2} \omega_e^{\mathrm{MF}}-\frac{1}{4}{\omega_{e}\chi_{e}}^{(\mathrm{MF})}),
\end{eqnarray}
where $E^{\mathrm{ MF{^+}}}$ and $E^{\mathrm{MF}}$ are the energies of the ion and the neutral systems (M = Ca,Sr,Ba) at the corresponding equilibrium bond lengths,  
$\omega_e^{\mathrm{MF}}$ is the vibrational frequency of MF$^{+}$ and $\omega_e^{\mathrm{MF}}$ is the vibrational frequency of MF (Table~\ref{constants}).

    \begin{table}[h]
        \centering
        \caption{Recommended theoretical IP of MF (M = Ca, Sr, Ba) [eV] with uncertainties.}\label{IPfinal}
        \begin{tabular}{llll}
        \hline
        IP & CaF&SrF&BaF  \\
        \hline 
        Adiabatic &5.821(2)&5.415(1)&4.807(3)\\  
        Adiabatic+ZPE&5.827(2)&5.420(1)&4.812(3)\\
        Experiment$^{a}$ &5.8279(2)&5.0(3)&4.80377(1)\\
        \hline
        \end{tabular}
        \begin{tablenotes}
        \footnotesize
        \item $^{a}$-CaF\cite{MurBerMer90},~SrF\cite{Hildenbrand1968} and BaF\cite{Jakubek1995}
        \end{tablenotes}
    \end{table} 

%Experimental value for SrF IP is from pure electron-impact measurement~\cite{Hildenbrand1968}. 

For CaF our results are in excellent agreement with experiment, showcasing the accuracy of our computational approach and the reliability of the scheme we use to assign uncertainties on our predictions. BaF, being the heaviest system of the three molecules discussed here, and thus most sensitive to the quality of its description, shows both the highest error bars and the largest discrepancy with experiment, while retaining very good agreement with the latter. In case of SrF, we obtained a much higher value that the available experiment. The rather large error bar on the measured value, combined with this discrepancy, could serve to motivate an accurate measurement of this and heavier monofluoride molecules such as RaF \cite{GarciaRuiz2020,udr21}.

% \begin{figure}[b]
%     \includegraphics{Figures/IP_atom_F.pdf}
%     \caption{Experimental ionization potentials of group 2 metal ions (blue) and their neutral monofluoride molecules (red).}
%     \label{fig:RaF_IP}
% \end{figure}

\section{Conclusions}

We performed calculations of the IPs of CaF, SrF, and BaF, using the
4-component (CCSD(T)) method and employing large
uncontracted basis sets with extrapolation to the CBS limit.
Breit, QED and higher order excitation corrections were
added a posteriori to the CCSD(T) values. An extensive analysis of the effect of the various computational parameters allowed us
to assign realistic error bars on our predictions. The calculated IPs of CaF and BaF are in excellent agreement with the experimental values, while for SrF we provide a new accurate prediction of the IP.

\section{Acknowledgments}
We thank the Center for Information Technology at the University of Groningen for their support and for providing access to the Peregrine high performance computing cluster and to the Hábrók high performance computing cluster. A.S. thank JSPS Overseas Challenge Program for Young Researchers, Grant No. 201880193, and JST Moonshot R\&D Grant No. JPMJMS2269. 
\bibliography{apssamp}% Produces the bibliography via BibTeM.

\end{document}